\documentclass[12pt]{article}
\usepackage{fullpage}
\usepackage{amssymb}
\usepackage{amsfonts}
\usepackage{latexsym}
\usepackage{times}
\usepackage{color}
\title{Correlation Clustering with a Fixed Number of Clusters}
\author{Ioannis Giotis \and Venkatesan Guruswami\thanks{Research supported in part by NSF grant CCF-0343672.}}

\date{Department of Computer Science and Engineering \\ University of
  Washington \\ Seattle, WA 98195. \\~ \\ {\tt
    \{giotis,venkat\}@cs.washington.edu}}
\newtheorem{theorem}{Theorem}
\newtheorem{lemma}[theorem]{Lemma}

\newcommand{\red}{ }

\def\diff{{\sf diff}}
\def\opt{{\sf OPT}}
\def\clusval{{\sf ClusVal}}
\def\clus{{\sf ClusMin}}
\def\clusmax{{\sf ClusMax}}
\def\val{{\sf val}}
\def\pval{{\sf pval}}
\def\cals{{\tilde{S}}}
\def\larg{{\sf Large}}
\def\smal{{\sf Small}}
\def\low{{\sf low}}
\newcommand{\cald}{{\cal D}}
\newcommand{\calf}{{\cal F}}
\newcommand{\cala}{{\cal A}}

\def\disagr{{\sf disagr}}
\newcommand{\eqdef}{\stackrel{{\rm def}}{=}}

\newcommand{\eps}{\varepsilon}
\renewcommand{\epsilon}{\varepsilon}
\newcommand{\mindisag}{{\sc MinDisAgree}}
\newcommand{\maxag}{{\sc MaxAgree}}
\newcommand{\maxagr}[1]{{\sc MaxAgree}$[#1]$}
\newcommand{\mindisagr}[1]{{\sc MinDisAgree}$[#1]$}
\newcommand{\maxcorr}{{\sc MaxCorr}}
\newcommand{\calc}{{\cal C}}

\newcommand{\etal}{{\rm et al.}}

\def\blackslug{\rule{2mm}{2mm}}
\def\QED{\hfill\blackslug}
\newcommand{\proof}{{\noindent {\bf \sl Proof\/}:\enspace}}

\newlength{\pgmtab}
\newenvironment{program}{%
\begin{tabbing}\hspace{0em}\=\hspace{0em}\=%
\hspace{\pgmtab}\=\hspace{\pgmtab}\=\hspace{\pgmtab}\=\hspace{\pgmtab}\=%
\hspace{\pgmtab}\=\hspace{\pgmtab}\=\hspace{\pgmtab}\=\hspace{\pgmtab}\=%
\+\+\kill}{\end{tabbing}}

\newcommand{\remove}[1]{}

\begin{document}

\maketitle
\thispagestyle{empty}

\begin{abstract}
  We continue the investigation of problems concerning {\em
    correlation clustering}~or {\em clustering with qualitative
    information}, which is a clustering formulation that has been
  studied recently~\cite{BBC,CGW,CW,AMMN}. The basic setup here is
  that we are given as input a complete graph on $n$ nodes (which
  correspond to nodes to be clustered) whose edges are labeled $+$
  (for similar pairs of items) and $-$ (for dissimilar pairs of
  items). Thus we have only as input qualitative information on
  similarity and no quantitative distance measure between items.  The
  quality of a clustering is measured in terms of its number of
  agreements, which is simply the number of edges it correctly
  classifies, that is the sum of number of $-$ edges whose endpoints
  it places in different clusters plus the number of $+$ edges both of
  whose endpoints it places within the same cluster.

In this paper, we study the problem of finding clusterings that maximize
    the number of agreements, and the complementary minimization
    version where we seek clusterings that minimize the number of
    disagreements. We focus on the situation when the number of
    clusters is stipulated to be a {\em small constant} $k$. Our main
    result is that for every $k$, there is a polynomial time
    approximation scheme for both maximizing agreements and minimizing
    disagreements. (The problems are NP-hard for every $k \ge 2$.)
The main technical work is for the minimization version, as the PTAS
    for maximizing agreements follows along the lines of the property
    tester for Max $k$-CUT from \cite{GGR}.

In contrast, when the number of clusters is not specified, the problem
    of minimizing disagreements was shown to be APX-hard~\cite{CGW},
    even though the maximization version admits a PTAS.

\end{abstract}
\newpage

\section{Introduction}

In this work, we continue the investigation of problems concerning an
appealing formulation of clustering called {\em correlation
  clustering}~ or {\em clustering using qualitative information} that
has been studied recently in several works, including \cite{BSY, SST, BBC,
  CGW, CW, AMMN}.  The basic setup here is to cluster a collection of
$n$ items given as input only qualitative information concerning
similarity between pairs of items; specifically for every pair of
items, we are given a (Boolean) label as to whether those items are
similar or dissimilar.  We are not provided with any quantitative
information on how different pairs of elements are, as is typically
assumed in most clustering formulations. These formulations take as
input a metric on the items and then aim to optimize some function of
the pairwise distances of the items within and across clusters. The
objective in our formulation is to produce a partitioning into
clusters that places similar objects in the same cluster and
dissimilar objects in different clusters, to the extent possible.

An obvious graph-theoretic formulation of the problem is the following: given a complete graph on $n$ nodes with each
edge labeled either ``$+$'' (similar) or ``$-$'' (dissimilar), find a partitioning of the vertices into clusters that
agrees as much as possible with the edge labels. The maximization version, call it \maxag\, seeks to maximize the
number of agreements: the number of $+$ edges inside clusters plus the number of $-$ edges across clusters. The
minimization version, denoted \mindisag, aims to minimize the number of disagreements: the number of $-$ edges within
clusters plus the number of $+$ edges between clusters.

In this paper, we study the above problems when the maximum number of
clusters that we are allowed to use is stipulated to be a fixed
constant $k$. We denote the variants of the above problems that have
this constraint as \maxagr{k} and \mindisagr{k}. We note that, unlike
most clustering formulations, the \maxag\ and \mindisag\ problems are
not trivialized if we do not specify the number of clusters $k$ as a
parameter --- whether the best clustering uses few or many clusters
is automatically dictated by the edge labels. However, the variants we
study are also interesting formulations, which are well-motivated in
settings where the number of clusters might be an external constraint
that has to be met, even if there are ``better'' clusterings (i.e.,
one with more agreements) with a different
number of clusters. Moreover, the existing algorithms for, say
\mindisag, cannot be modified in any easy way to output a quality
solution with at most $k$ clusters. Therefore $k$-clustering
variants pose new, non-trivial challenges that require different
techniques for their solutions.

In the above description, we have assumed that every pair of items is
labeled as $+$ or $-$ in the input.  In a more general variant,
intended to capture situations where the classifier providing the
input might be unable to label certain pairs of elements are similar
or dissimilar, the input is an arbitrary graph $G$ together with $\pm$
labels on its edges. We can again study the above problems \maxagr{k}
(resp. \mindisagr{k}) with the objective being to maximize (resp.
minimize) the number of agreements (resp. disagreements) on edges of
$E$ (that is, we do not count non-edges of $G$ as either agreements or
disagreements). In situations where we study this more general
variant, we will refer to these problems as \maxagr{k} on {\em
  general} graphs and \mindisagr{k} on general graphs. When we don't
qualify with the phrase ``on general graphs'', we will always mean the
problems on complete graphs.

Our main result in this paper is a polynomial time approximation
scheme (PTAS) for \maxagr{k} as well as \mindisagr{k} for $k \ge 2$.
We now discuss prior work on these problems, followed by a more
detailed description of results in this paper.

\subsection{Previous and related work}
The above problem seems to have been first considered by Ben-Dor
\etal~\cite{BSY} motivated by some computational biology questions.
Later, Shamir \etal~\cite{SST} studied the computational complexity of
the problem and showed that \maxag\ (and hence also \mindisag), as
well as \maxagr{k} (and hence also \mindisagr{k}) for each $k \ge 2$
is NP-hard. They, however, used the term ``Cluster Editing'' to refer
to this problem.

Partially motivated by some machine learning problems concerning
document classification, Bansal, Blum, and Chawla~\cite{BBC} also
independently formulated and considered this problem. In particular,
they initiated the study of approximate solutions to \mindisag\ and
\maxag , and presented a PTAS for \maxag\ and a constant factor
approximation algorithm for \mindisag\ (the approximation guarantee
was a rather large constant, though).  They also noted a simple factor
$3$ approximation algorithm for \mindisagr{2}.  Charikar, Guruswami
and Wirth~\cite{CGW} proved that \mindisag\ is APX-hard, and thus one
cannot expect a PTAS for the minimization problem similar to the PTAS
for \maxag. They also gave a factor $4$ approximation algorithm for
\mindisag\ by rounding a natural linear programming relaxation using
the region growing technique.

The problems on general graphs have also received attention.
It is known that both \maxag\ and \mindisag\ are APX-hard~\cite{BBC,CGW}.
Using a connection to minimum multicut, several
groups~\cite{CGW,nickle,dotan} presented an $O(\log n)$ approximation
algorithm for \mindisag. In fact, it was noted in \cite{dotan} that
the problem is as hard to approximate as minimum multicut (and so
this $\log n$ factor seems very hard to improve). For the maximization
version, algorithms with performance ratio better than $0.766$ are
known for \maxag~\cite{CGW,swamy}. The latter work by Swamy~\cite{swamy}
shows that a factor $0.7666$ approximation can also be achieved when
the number of clusters is specified (i.e., for \maxagr{k} for $k \ge
2$).

Another problem that has been considered, let us call it \maxcorr, is
that of maximizing correlation, defined to be the difference between
the number of agreements and disagreements. A factor $O(\log n)$
approximation for \maxcorr\ on complete graphs is presented in
\cite{CW}. Recently \cite{AMMN} showed an integrality gap of
$\Omega(\log n)$ for the underlying semidefinite program used in
\cite{CW}.  They also prove that an approximation of $O(\log
\theta(\overline{G}))$ can be achieved on general graphs $G$, where
$\theta(\cdot)$ is the Lov\'{a}sz Theta Function.

\subsection{Our results}

The only previous approximation for \mindisagr{k} was a factor $3$ approximation algorithm for the case
$k=2$~\cite{BBC}. The problems were shown to be NP-hard for every $k \ge 2$ in \cite{SST} \red{using a rather
complicated reduction. In this paper, we will provide a much simpler NP-hardness proof and} prove that both \maxagr{k}
and \mindisagr{k} admit a polynomial time approximation scheme for every $k \ge 2$.\footnote{Our
  approximation schemes will be randomized and deliver a solution with
  the claimed approximation guarantee with high probability. For
  simplicity, we do not explicitly mention this from now on.} These
approximation schemes are presented in Section~\ref{sec:maxagr} and
\ref{sec:mindisagr} respectively.  The existence of a PTAS for
\mindisagr{k} is perhaps surprising in light of the APX-hardness of
\mindisag\ when the number of clusters is not specified to be a
constant (recall that the maximization version {\em does} admit a PTAS
even when $k$ is not specified).

It is often the case that minimization versions of problems are harder
to solve compared to their complentary maximization versions. The
APX-hardness of \mindisag\ despite the existence of a PTAS for \maxag\
is a notable example.  The difficulty in these cases is when the
optimum value of the minimization version is very small, since then
even a PTAS for the complementary maximization problem need not
provide a good approximation for the minimization problem.  In this
work, we first give a PTAS for \maxagr{k}. This algorithm uses random
sampling and follows closely along the lines of the property testing
algorithm for Max $k$-Cut due to \cite{GGR}. We then develop a PTAS
for \mindisagr{k}, which is our main result. This requires more work
and the algorithm returns the better of two solutions, one of which is
obtained using the PTAS for \maxagr{k}.

The difficulty in getting a PTAS for the minimization version is
similar to that faced in the problem of Min $k$-sum clustering, which
has the complementary objective function to Metric Max $k$-Cut. We
remark that while an elegant PTAS for Metric Max $k$-Cut due to de la
Vega and Kenyon~\cite{vega-kenyon} has been known for several years,
only recently has a PTAS for Min $k$-sum clustering been
obtained~\cite{dVKKR}.  We note that the case of Min $2$-sum
clustering though was solved in \cite{indyk} soon after the Metric Max
Cut algorithm of \cite{vega-kenyon}, but the case $k > 2$ appeared
harder. Similarly to this, for \mindisagr{k}, we are able to quite
easily give a PTAS for the $2$-clustering version using the algorithm
for \maxagr{2}, but we have to work harder for the case of $k > 2$
clusters. Some of the difficulty that surfaces when $k > 2$ is
detailed in Section~\ref{subsec:k-diff}.

In Section~\ref{sec:general-ghs}, we also note some results on the
complexity of \maxagr{k} and \mindisagr{k} on general graphs --- these
are easy consequences of connections to problems like Max CUT and
graph colorability.

\medskip Our work seems to nicely complete the understanding of the
complexity of problems related to correlation clustering. Our
algorithms not only achieve excellent approximation guarantees but are
also sampling-based and are thus simple, combinatorial, and quite easy
to implement.  To compare with the situation for the case when $k$ is
not specified, the algorithm for \mindisag\ in \cite{BBC} achieves a
very large approximation factor. On the other hand, the algorithm in
\cite{CGW} achieves a good factor of $4$ but needs to solve a linear
programming relaxation.  In fact it could well be that on some
instances we can find a better solution compared to what the algorithm of
\cite{CGW} produces even when $k$ isn't specified by trying our
algorithm for some small values of $k$.


\section{NP-hardness of \mindisag\ and \maxag}
In this section we show that the exact versions of problems we are trying to solve are NP-hard. An NP-hardness result
for \maxag\  on complete graphs was shown in \cite{BBC}; however their reduction crucially relies on the number of
clusters growing with the input size, and thus does not yield any hardness when the number of clusters is a fixed
constant $k$. It was shown by Shamir, Sharan, and Tsur~\cite{SST}, using a rather complicated reduction, that these
problems are NP-hard for each fixed number $k \ge 2$ of clusters. We will provide a short and intuitive proof that
\mindisagr{k} and \maxagr{k} are NP-hard.

Clearly it suffices to establish the NP-hardness of \mindisagr{k} since \maxagr{k} can be easily reduced on a
complimentary graph. We will first establish NP-hardness for $k=2$, the case for general $k$ will follow by a simple
``padding'' with $(k-2)$ large collection of nodes with $+$ edges between nodes in each collection and $-$ edges to
everywhere else.

\begin{theorem}
\label{thm:np-hardness-min2} \mindisagr{2} on complete graphs is NP-hard.
\end{theorem}

\proof We know that Graph Min Bisection, namely partitioning the vertex set of a graph into two equal halves so that
the number of edges connecting vertices in different halves is minimized, is NP-hard. From an instance $G$ of Min
Bisection with $n$(even) vertices we obtain a complete graph $G'$ using the following polynomial time construction.

Start with $G$ and label all existing edges of $G$ as $+$ edges in $G'$ and non-existing edges as $-$ edges. For each
vertex $v$ create an additional set of $n$ vertices. Let's call these vertices together with $v$, a ``group'' $V_{v}$.
Connect with $+$ edges all pairs of vertices within $V_{v}$. All other edges with one endpoint in $V_{v}$ as labeled as
$-$ edges (except those already labeled).

We will now show that any 2-clustering of $G'$ with the minimum number of disagreements, has 2 clusters of equal size
with all vertices of any group in the same cluster. Consider some optimal 2-clustering $W$ with 2 clusters $W_{1}$ and
$W_{2}$ such that $|W_{1}|\neq |W_{2}|$ or not all vertices of some group are in the same cluster. Pick some group
$V_{v}$ such that not all its vertices are assigned in the same cluster. If such a group cannot be found, pick a group
$V_{v}$ from the larger cluster. Place all the vertices of the group in the same cluster obtaining $W'$ such that $|
|W'_{1}| - |W'_{2}||$ is minimized.

Let's assume that $V_{v}^{1}$ vertices of group $V_{v}$ were in $W_{1}$ and $V_{v}^{2}$ in $W_{2}$. Wlog, let's assume
that $W'$ is obtained by moving the $V_{v}^{1}$ group vertices in cluster $W_{2}$.
\[ W'_{1}=W_{1}\setminus V_{v}^{1}, W'_{2}=W_{2}\cup V_{v}^{1} \]

We now observe the following facts about the difference in the number of disagreements between $W'$ and $W$.
\begin{itemize}
\item  Clearly the number of disagreements between vertices not in $V_{v}$ and between one vertex in $V_{v}^{2}$ with one in $W'_{1}$  remains the same.
\item The number of disagreements is decreased by $|V_{v}^{1}|\cdot |V_{v}^{2}|$
based on the fact that all edges within $V_{v}$ are $+$ edges.
\item It is also decreased by at least $ |V_{v}^{1}| \cdot | W'_{1}| - (n-1)$ based on the fact that all
but at most $n-1$ edges connecting vertices of $V_{v}$ to the rest of the graph are $-$ edges.
\item The number of disagreements increases at most $|V_{v}^{1}|\cdot |W_{2}\setminus V_{v}^{2}|$ because (possibly)
all of the vertices in $V_{v}^{1}$ are connected with $-$ edges with vertices in $W_{2}$ outside their group.
\end{itemize}

Overall, the difference in the number of disagreements is at most $|V_{v}^{1}|\cdot |W_{2}\setminus V_{v}^{2}| -
|V_{v}^{1}|\cdot |V_{v}^{2}| - |V_{v}^{1}| \cdot | W'_{1}| + (n-1)$. Notice that since $| |W'_{1}| - |W'_{2}||$ was
minimized it must be the case that $| W'_{1}|\geq |W_{2}\setminus V_{v}^{2}|$. Moreover since a group has an odd number
of vertices and the total number of vertices of $G'$ is even, it follows that $|W'_{1}|\neq |W_{2}\setminus V_{v}^{2}|$
and $| W'_{1}|-|W_{2}\setminus V_{v}^{2}|\geq 1$. Therefore the total number of disagreements increases at most $(n-1)-
|V_{v}^{1}|\cdot (|V_{v}^{2}|+1)$. Since $|V_{v}^{1}|+|V_{v}^{2}|=n+1$ and $V_{v}^{1}$ cannot be empty, it follows that
$|V_{v}^{1}|\cdot (|V_{v}^{2}|+1)\geq n$ and the number of disagreements strictly decreases contradicting the
optimality of $W$.

Therefore the optimal solution to the \mindisagr{2} instance has 2 clusters of equal size and all vertices of any group
are contained in a single cluster. It is now trivial to see that an optimal solution to the Min Bisection problem can
be easily derived from the \mindisagr{2} solution which completes the reduction. \QED

We are now able to easily derive the following NP-hardness result.
\begin{theorem}
\label{thm:np-hardness} For every $k \ge 2$, the problems \maxagr{k} and \mindisagr{k} on complete graphs are NP-hard.
\end{theorem}

\proof Consider an instance of the \mindisagr{2} problem on a graph $G$ with $n$ vertices. Create a graph $G'$ by
adding to $G$, $k-2$ ``groups'' of $n+1$ vertices each. All edges within a group are marked as $+$ edges, while the
remaining edges are marked as $-$ edges.

Consider now a $k$-clustering of $G'$ such that the number of disagreements is minimized. It is easy to see that all
the vertices of a group must make up one cluster. Also observe that any of the original vertices cannot end up in one
group's cluster since that would induce $n+1$ disagreements, stricly more than it could possibly induce in any of the 2
remaining clusters. Therefore the 2 non-group clusters are an optimal 2-clustering of $G$. The theorem easily follows.
\QED

\section{PTAS for maximizing agreement with $k$ clusters}
\label{sec:maxagr} \red{In this section we will present a PTAS for \maxagr{k} for every fixed constant $k$. Our
algorithm follows closely the PTAS for Max $k$-CUT by Goldreich et al.\cite{GGR}. In the next section, we will present
our main result, namely a PTAS for \mindisagr{k}, using the PTAS for \maxagr{k} together with additional
ideas.}\footnote{This
  is also similar in spirit, for example, to the PTAS for Min 2-sum clustering based on the PTAS for Metric Max
  CUT~\cite{indyk,vega-kenyon}.}

\begin{theorem}
\label{maxagr1}
  For every $k\ge 2$, there is a polynomial time approximation scheme
  for \maxagr{k}.
\end{theorem}
\proof We first note that for every $k \ge 2$, and every instance of
\maxagr{k}, the optimum number
$\opt$ of
agreements is at least $n^2/16$.
Let $n_+$ be the number of positive edges, and $n_- = {n \choose 2} -
n_+$ be the number of negative edges. By placing all vertices in a
single cluster, we get $n_+$ agreements. By placing vertices randomly
in one of $k$ clusters, we get an expected $(1-1/k)n_-$ agreements
just on the negative edges. Therefore $\opt \ge \max \{n_+ ,
(1-1/k)n_-\} \ge (1-1/k){n \choose 2}/2 \ge n^2/16$. The proof now
follows from Theorem \ref{maxagr2} which guarantees a solution within
additive $\eps n^2$ of $\opt$ for arbitrary $\eps > 0$. \QED

\begin{theorem}\label{maxagr2}
  On input $\epsilon$, $\delta$ and a labeling ${\cal L}$ of the edges
  of a complete graph $G$ with $n$ vertices, with probability at least
  $1-\delta$, algorithm \emph{MaxAg} outputs a $k$-clustering of the
  graph such that the number of agreements induced by this
  $k$-clustering is at least $\opt- \epsilon n^{2}/2$, where $\opt$ is
  the optimal number of agreements induced by any $k$-clustering of
  $G$. The running time of the algorithm is $\frac{n}{\eps}
  k^{O(\eps^{-2} \log k \log (1/\eps\delta))}$.
\end{theorem}

The proof of this theorem is presented in Section \ref{MaxProof}, and
we now proceed to describe the algorithm.

\medskip
\noindent \emph{Algorithm} {\bf MaxAg}$(k,\eps)$:

\smallskip
\noindent \underline{Input}: A labeling ${\cal L}: {n \choose 2} \rightarrow \{+,-\}$ of the edges of the complete
graph on vertex set $V$. 

\noindent \underline{Output}: A $k$-clustering of the graph, i.e., a partition of $V$ into (at most) $k$ parts
$V_1,V_2,\dots,V_k$.

\begin{program}
1. Construct an arbitrary partition of the graph $(V^{1},V^{2},\ldots,V^{m}),m=\lceil\frac{4}{\epsilon}\rceil$.\\
2. For $i=1\ldots m$, choose uniformly at random with replacement
  from $V\setminus V^{i}$,\\
 \hspace*{0.5cm} a subset $S^{i}$ of size $r=\Theta \left( \frac{1}{\epsilon^{2}} \log\frac{1}{\epsilon \delta} \log k \right) $.\\
3. For $i=1\ldots m$ do the following\+ \\
   (a) For each clustering of $S^{i}$ into $(S^{i}_{1},\ldots ,S^{i}_{k})$ do the following\+ \\
      (i)   For each vertex $v\in V^{i}$ do\+ \\
          (1)  For $j=1\dots k$, let $\beta_{j}(v) = |\Gamma^{+}(v)\cap S^{i}_{j}| + \sum_{l\neq j}|\Gamma^{-}(v)\cap
          S^{i}_{l}|$.\\
          (2)  Place $v$ in $W^{i}_{j}$ for which $\beta_{j}(v)$ is maximized.\- \\
      (ii) \red{If the clustering on the subgraph induced by the edges between $V^{j}$ and $S^{i}$ }\\
            \hspace*{0.5cm} \red{has more agreements than the currently stored one, store this clustering as $W^{i}\equiv (W^{i}_{1}, \ldots , W^{i}_{k})$ }\-\- \\
4. For $j=1\ldots k$, let $W_{j}\equiv \cup_{i}W^{i}_{j}$. Output clustering $(W_{1},\ldots, W_{k})$.
\end{program}

\subsection{Overview}
Our algorithm is given a complete graph $G(V,E)$ on $n$ vertices. All
the edges are marked as $+$ or $-$, denoting whether adjacent vertices
are on agreement or disagreement respectively. For a vertex $v$, let
$\Gamma^{+}(v)$ be the set of vertices adjacent to $v$ via $+$ edges,
and $\Gamma^{-}(v)$ the set of vertices adjacent to $v$ via $-$ edges.

The algorithm works in $m=O(1/\epsilon$) steps. At each step we are
placing $\Theta(\epsilon n)$ vertices into clusters. We will show that
with constant probability our choices of $S_{i}$'s will allow us to
place the vertices in such a way that the decrease in the number of
agreements with respect to an optimal clustering is
$O(\epsilon^{2}n^{2})$ per step, thus the algorithm outputs a solution
that has $O(\epsilon n^{2})$ less agreements than any optimal
solution.

\subsection{\label{MaxProof}Performance analysis of {\bf \textit{MaxAg}}$(k,\epsilon)$ algorithm}
Consider an arbitrary optimal $k$-clustering of the graph $D\equiv(D_{1},\ldots , D_{k})$.  \red{We consider the
subsets of each cluster over our partition of vertices, defined as}
\begin{eqnarray*}
\mbox{for } j=1,\ldots k,\ D^{i}_{j} & \equiv & D_{j} \cap V^{i} \\
D^{i} & \equiv & (D^{i}_{1},\ldots , D^{i}_{k})
\end{eqnarray*}
We will now define a sequence of \emph{hybrid} clusterings, such that hybrid clustering $H^{i}$, for $i=1,2,\dots,m+1$,
consists of the vertices as clustered by our algorithm up to \red{(not including)} the $i$'th step and the rest of the
vertices as clustered by $D$.
\begin{eqnarray*}
H^{i} & \equiv & (H^{i}_{1},\ldots ,H^{i}_{k}) \\
\mathcal{H}^{i} & \equiv & (\mathcal{H}^{i}_{1},\ldots ,\mathcal{H}^{i}_{k}) \\
\mbox{for } j=1,\ldots k,\ H^{i}_{j} & \equiv & (\cup_{l=1}^{i-1} W^{l}_{j}) \cup (\cup_{l=i}^{m} D^{l}_{\red{j}}) \\
\mbox{for } j=1,\ldots k,\ \mathcal{H}^{i}_{j} & \equiv & H^{i}_{j}\setminus V^{i}
\end{eqnarray*}
Although we will go through all possible clusterings of $S^{i}$, for
the rest of the analysis consider the particular clustering that
matches the hybrid clusterings,
\[ \mbox{for } j=1\ldots k,\ S^{i}_{j}\equiv S^{i} \cap \mathcal{H}^{i}_{j} \]

\red{The following theorem captures the fact that our random sample with high probability gives us a good estimate on
the number of agreements towards each cluster for most of the vertices considered.}
\begin{lemma}\label{chernoff}
  For $i=1\ldots m$, with probability at least $1- (\delta/4m)$ on the
  choice of $S^{i}$, for all but at most an $\epsilon / 8$ fraction of
  the vertices $v\in V^{i}$, the following holds
\begin{eqnarray}
\label{representative} \mbox{for } j=1,\ldots k,\ \left| \frac{ |\Gamma^{+}(v)\cap S^{i}_{j}| }{r} -
\frac{|\Gamma^{+}(v)\cap \mathcal{H}^{i}_{j}| }{|V\setminus V^{i}|}
\right| \leq \frac{\epsilon}{32} \ .
\end{eqnarray}
(Note that if (\ref{representative}) above holds, then it also holds
with $\Gamma^-(v)$ in place of $\Gamma^+(v)$.)
\end{lemma}
\proof Consider an arbitrary vertex $v\in V^{i}$ and the randomly
chosen set $S^{i}=\{u_{1},\ldots,u_{r}\}$. For each
$j\in\{1,\ldots,k\}$, we define the random variables
\[ \mbox{for }  l=1,\ldots r,\ \alpha^{l}_{j}=\left\{ \begin{array}{ll}
1, & \mbox{if } u_{l}\in \Gamma^{+}(v)\cap S^{i}_{j}; \\
0, & otherwise.
\end{array}\right.
\]

Clearly $ \sum_{l=1}^{r}\alpha^{l}_{j} = |\Gamma^{+}(v)\cap S^{i}_{j}|
$ and $Pr[\alpha^{l}_{j} =1] = \frac{|\Gamma^{+}(v)\cap
  \mathcal{H}^{i}_{j}| }{|V\setminus V^{i}|}$.

Using an additive Chernoff bound we get that
\begin{eqnarray} Pr\left[ \left| \frac{|\Gamma^{+}(v)\cap S^{i}_{j}|}{r} - \frac{|\Gamma^{+}(v)\cap
\mathcal{H}^{i}_{j}|}{|V\setminus V^{i}|} \right| > \frac{\epsilon}{32} \right] < 2\cdot \exp(-2( \frac{\epsilon}{32}
)^{2} r) < \frac{\epsilon \delta}{32 mk} \end{eqnarray}

\red{Defining a random variable to count the number of vertices not satisfying inequality(\ref{representative}) and
using Markov's inequality we get that for that particular $j$, inequality(\ref{representative}) holds for all but a
fraction $\epsilon /8$ of vertices $v\in V^{i}$, with probability at least $1-(\delta /4mk)$. Using a probability union
bound the lemma easily follows.} \QED

\smallskip
We define ${\sf agree}(A)$ to be equal to the number of agreements induced by $k$-clustering $A$. Now consider the
placement of $V^{i}$ vertices in clusters $W^{i}_{1},\ldots, W^{i}_{k}$ as performed by the algorithm during step $i$.
We will examine the number of agreements compared to the placement of the same vertices under $H^{i}$ \red{(placement
under the optimal clustering)}, more specifically we will bound the difference in the number of agreements induced by
placing vertices differently than $H^{i}$. The following lemma formalizes this concept.

\begin{lemma}\label{maxlem1}
For $i=0,\ldots m$, we have ${\sf agree}(H^{i+1}) \geq {\sf agree}(D)
  - i\cdot \frac{1}{8}\epsilon^{2}n^{2} $
\end{lemma}
\proof Observe that $H^{\red{1}}\equiv D$ and $H^{m+1}\equiv W$. The only vertices placed differently between $H^{i+1}$
and $H^{i}$ are the vertices in $V^{i}$. Suppose that our algorithm places $v\in V^{i}$ in cluster $x$ and $v$ is
placed in cluster $x'$ under $H^{i}$. For each vertex $v$ the number of agreements towards clusters other than $x,x'$
remains the same, therefore we will focus on the number of agreements towards these two clusters and the number of
agreements within $V^{i}$.

The number of agreements we could lose by thus misplacing $v$ is
\begin{eqnarray*}
\diff_{xx'}(v)=|\Gamma^{+}(v)\cap \mathcal{H}^{i}_{x'}| - |\Gamma^{+}(v)\cap \mathcal{H}^{i}_{x}| + |\Gamma^{-}(v)\cap
\mathcal{H}^{i}_{x}| - |\Gamma^{-}(v)\cap \mathcal{H}^{i}_{x'}|
\end{eqnarray*}

\red{Since our algorithm chose cluster $x$, by construction
\begin{equation}\label{eq:diff1}
|\Gamma^{+}(v)\cap S^{i}_{x}|+|\Gamma^{-}(v)\cap S^{i}_{x'}| \geq |\Gamma^{+}(v)\cap S^{i}_{x'}|+|\Gamma^{-}(v)\cap
S^{i}_{x}|
\end{equation}}

\red{If inequality (\ref{representative}) holds for vertex $v$, using it for $\Gamma^{+}(v)$ and $\Gamma^{-}(v)$ in
both clusters $x$,$x'$, we obtain bounds on the difference of agreements between our random sample's clusters
$S^{i}_{x},S^{i}_{x'}$ and the hybrid clusters $\mathcal{H}^{i}_{x},\mathcal{H}^{i}_{x'}$. Combining with inequality
(\ref{eq:diff1}) we get that $\diff_{xx'}(v)$ is at most $\frac{1}{8}\epsilon n$. Therefore the total decrease in the
number of agreements by this type of vertices is at most $\frac{1}{8}\epsilon n|V^{i}|\leq \frac{1}{8}\epsilon
\frac{n^{2}}{m}$.}

\red{By Lemma \ref{chernoff} there are at most $(\epsilon /8)|V^{i}|$ vertices in $V^{i}$ for which inequality
(\ref{representative}) doesn't hold. The total number of agreements originating from these vertices is at most
$\frac{1}{8}\epsilon |V^{i}|n \leq \frac{1}{8}\epsilon \frac{n^{2}}{m}$. Finally, the total number of agreements from
within $V_{i}$ is at most $|V^{i}|^{2} \leq \frac{1}{4}\epsilon \frac{n^{2}}{m} $.}

Overall the number of agreements that we could lose in one step of the
algorithm is at most $\frac{1}{2}\epsilon \frac{n^{2}}{m}\leq
\frac{1}{8}\epsilon^{2}n^{2}$. The lemma follows by induction. \QED

\smallskip
The approximation guarantee of Theorem \ref{maxagr2} easily follows from Lemma \ref{maxlem1}. \red{At each step of the
algorithm we need to go over all $k^{r}$ $k$-clusterings of our random sample $S^{i}$ and there are $O(n/\epsilon)$
steps. All other operations within a step can be easily implemented to run in constant time and the running time bound
of our algorithm follows as well.} \QED

\section{PTAS for minimizing disagreements with $k$ clusters}
\label{sec:mindisagr}
This section is devoted to the proof of the following theorem, which
is our main result in this paper.
\begin{theorem}[Main]
\label{thm:min-ptas}
For every $k\ge 2$, there is a PTAS for \mindisagr{k}.
\end{theorem}
The algorithm for \mindisagr{k} will use the
approximation scheme for \maxagr{k} as a
subroutine. The latter already provides a very good
approximation for the number of disagreements unless this number is
very small. So in the analysis, the main work is for the case when the
optimum clustering is right on most of the edges.
\subsection{Idea behind the algorithm}
\label{subsec:k-diff}
The case of $2$-clusters turns out to be lot simpler and we use it to first illustrate the basic idea. By the PTAS for
maximization, we only need to focus on the case when the optimum clustering has only $\opt = \gamma n^2$ disagreements
for some small $\gamma > 0$.  We draw a random sample $S$ and try all partitions of it, and focus on the run when we
guess the right partition $S = S_1 \cup S_2$, namely the way some fixed optimal clustering $\cald$ partitions $S$.
Since the optimum has very large number of agreements, each node in a set $A$ of size at least $(1-O(\gamma))n$ will
have a clear choice of which side they prefer to be on and we can find this out with high probability based on edges
into $S$.  Therefore, we can find a clustering which agrees with $\cald$ on a set $A$ of at least $1-O(\gamma)$
fraction of the nodes. We can then go through this clustering and for each node in parallel switch it to the other side
if that improves the solution to produce the final clustering. Nodes in $A$ won't get switched and will remain
clustered exactly as in the optimum $\cald$.  The number of extra disagreements compared to $\cald$ on edges amongst
nodes in $V - A$ is obviously at most the number of those edges which is $O(\gamma^2 n^2)$.  For edges connecting a
node $u \in V - A$ to nodes in $A$, since we placed $u$ on the ``better'' side, and $A$ is placed exactly as in $\cald$
in the final clustering, we can have at most $O(\gamma n)$ extra disagreements \red{per node compared to $\cald$,
simply the error introduced by the edges towards the misplaced nodes in $V - A$}. Therefore we get a clustering with at
most $\opt + O(\gamma^2 n^2) \le (1+O(\gamma)) \opt$ disagreements.

Our algorithm for $k$-clustering for $k > 2$ uses a similar high-level
approach, but is more complicated. The main thing which breaks down
compared to the $k=2$ case is the following. For two clusters, if
$\cald$ has agreements on a large, i.e. $(1-O(\gamma))$, fraction of
edges incident on a node $u$ (i.e. if $u \in A$ in the above
notation), then we are guaranteed to place $u$ exactly as in $\cald$
based on the sample $S$ (when we guess its correct clustering), since
the other option will have much poorer agreement.  This is not the
case when $k > 2$, and one can get a large number of agreements by
placing a node in say one of two possible clusters. It therefore does
not seem possible to argue that each node in $A$ is correctly placed,
and then to use this to finish off the clustering.

However, what we {\em can} show is that nodes in $A$ that are incorrectly placed, call this set $B$, must be in small
clusters of $\cald$, and thus are few (at most $O(n/k)$) in number. Moreover, every node in $A$ that falls in one of
the large clusters that we produce, is guaranteed to be correctly placed. (These facts are the content of
Lemma~\ref{lem:large-clus}.) The nodes in $B$ still need to be clustered, and since they could be of size
$\Omega(n/k)$, even a small number of mistakes per node in clustering them is more than we can afford. We get around
this predicament by noting that nodes in $B$ and $A - B$ are in different sets of clusters in $\cald$. It follows that
we can cluster $B$ recursively in new clusters (and we are making progress because $B$ is clustered using fewer than
$k$ clusters).  The actual algorithm must also deal with nodes outside $A$, and in particular decide which of these
nodes are recursively clustered along with $B$. With this intuition in place, we now proceed to the formal
specification of the algorithm.

\subsection{Algorithm for $k$-clustering to minimize disagreements}

The following is the algorithm that gives a factor $(1+\eps)$
approximation for \mindisagr{k}. We will use a small enough absolute
constant $c_1$ in the algorithm; the choice $c_1 =1/20$ will work.

\smallskip
\noindent \emph{Algorithm} {\bf MinDisAg}$(k,\eps)$:

\smallskip
\noindent \underline{Input}: A labeling ${\cal L} : {n \choose 2}
\rightarrow \{+,-\}$ of the edges of the complete graph on vertex set $V = \{1,2,\dots,n\}$

\noindent \underline{Output}: A $k$-clustering of the graph, i.e., a
partition of $V$ into (at most) $k$ parts $V_1,V_2,\dots,V_k$.

\begin{program}
0. If $k=1$, return the obvious $1$-clustering.\\
1. Run the PTAS for \maxagr{k} from previous section on input
  ${\cal L}$ with accuracy  $\frac{\eps^2 c_1^2}{32 k^4}$. \\
\hspace*{0.5cm} Let $\clusmax$ be the $k$-clustering returned.\\
2. Set $\beta = \frac{c_1 \eps}{16 k^2}$. Pick a sample $S \subseteq V$ by
drawing $\frac{5 \log n}{\beta^2}$
  vertices u.a.r with replacement.\\

3. $\clusval \leftarrow 0$; /* Keeps track of value of best clustering
  found so far*/ \\
4. For each partition $\cals$ of $S$ as $S_1 \cup S_2 \cup \cdots \cup
  S_k$,
  perform the following steps:\+ \\
(a) Initialize the clusters $C_i = S_i$ for $1\le i \le k$.\\
(b) For each $u \in V - S$\+ \\
(i) For each $i=1,2,\dots,k$, compute $\pval^\cals(u,i)$, defined to be
  $1/|S|$ times  the number
  of \\\hspace*{0.5cm} agreements on edges connecting $u$ to nodes in $S$ if $u$ is
  placed in cluster $i$ along with $S_i$.\\
(ii) Let $j_u = \mbox{arg max}_i
  \pval^\cals(u,i)$, and $\val^\cals(u) \eqdef \pval^\cals(u,j_u)$. \\
(iii) Place $u$ in cluster $C_{j_u}$, i.e., $C_{j_u} \leftarrow C_{j_u}  \cup
  \{u\}$. \- \\
(c) Compute the set of large and small clusters as \\
\hspace*{0.6cm}
$\larg = \{ j \mid
  1 \le j \le k, ~|C_j| \ge \frac{n}{2k}\}$,
and $\smal = \{1,2,\dots,k\} - \larg$. \\
\hspace*{0.6cm}Let $l =|\larg|$ and $s
  = k - l =|\smal|$. /* Note that $s < k$. */ \\

(d) Cluster $W \eqdef \bigcup_{j \in \smal} C_j$ into $s$
clusters using recursive call to algorithm {\bf
  MinDisAg}$(s,\eps/10)$. \\
\hspace*{0.6cm}
Let the clustering output by the recursive call be $W = W'_1 \cup W'_2
\cup \cdots \cup W'_s$ \\ \hspace*{0.6cm} (where some of the $W'_i$'s
may be empty) \\
(e) Let ${\cal C}$ be the clustering comprising of the $k$ clusters
$\{C_j\}_{j \in \larg}$ and $\{W'_i\}_{1 \le i \le s}$. \\
\hspace*{0.6cm} If the number of agreements of ${\cal C}$ is at least
$\clusval$, update $\clusval$ to this value, and \\
\hspace*{0.6cm} update  $\clus \leftarrow
{\cal C}$.\-\- \\
5. Output the better of the two clusterings $\clusmax$ and $\clus$.
\end{program}

\subsection{Performance analysis of the algorithm}
We now analyze the approximation guarantee of the above algorithm.  We
need some notation.  Let $\cala = A_1 \cup A_2 \cup \cdots A_k$ be any
$k$-clustering of the nodes in $V$.  Define the function $\val^{\cala}
: V \rightarrow [0,1]$ as follows: $\val^{\cala}(u)$ equals the
fraction of edges incident upon node $u$ whose labels agree with
clustering $\cala$ (i.e., we count negative edges that are cut by
$\cala$ and positive edges that lie within the same $A_i$ for some
$i$). Also define $\disagr(\cala)$ to be the number of disagreements
of $\cala$ w.r.t labeling $L$. (Clearly $\disagr(\cala) =
\frac{n-1}{2}\sum_{u \in V} (1-\val^\cala(u))$.) For a node $u \in V$
and $1 \le i \le k$, let $\cala^{(u,i)}$ denote the clustering
obtained from $\cala$ by moving $u$ to $A_i$ and leaving all other
nodes untouched.  We define the function $\pval^\cala : V \times
\{1,2,\dots,k\} \rightarrow [0,1]$ as follows: $\pval^\cala(u,i)$
equals the fraction of edges incident upon $u$ that agree with the
clustering $\cala^{(u,i)}$.

In the following, we fix $\cald$ to be any optimal $k$-clustering that
partitions $V$ as $V = D_1 \cup D_2 \cup \cdots \cup D_k$. Let
$\gamma$ be defined to be so that
$\disagr(\cald)/n^2$, i.e., the clustering $\cald$ has
$\gamma n^2$ disagreements w.r.t. the input labeling $L$.

Call a sample $S$ of nodes, each drawn
uniformly at random with replacement, to be $\alpha$-good if the nodes in $S$
are distinct\footnote{Note that in the algorithm we draw elements of
  the sample with replacement, but for the analysis, we
  can pretend that $S$ consists of distinct elements, since this
  happens with high probability.}
and for each $u \in V$ and $i \in \{1,2,\dots,k\}$,
\begin{equation}
\label{eq:goodness}
| \pval^\cals(u,i) - \pval^\cald(u,i) | \le \alpha
\end{equation}
for the partition $\cals$ of $S$ as $\cup_{i=1}^k S_i$ with $S_i = S
\cap D_i$ (where $\pval^\cals(\cdot,\cdot)$ is as defined in the
algorithm). The following lemma follows by a standard Chernoff and
union bound argument similar to Lemma~\ref{chernoff}.\footnote{Since
  our sample size is $\Omega(\log n)$ as opposed to $O(1)$ that was used in
  Lemma~\ref{chernoff}, we can actually ensure (\ref{eq:goodness}) holds for
  {\em every} vertex w.h.p.}
\begin{lemma}
\label{lem:sample-goodness} The sample $S$ picked in Step 2 is $\beta$-good with high probability ~(at least
$1-O(1/\sqrt{n})$).
\end{lemma}

Therefore, in what follows we assume that the sample $S$ is
$\beta$-good. In the rest of the discussion, we focus on the run of
the algorithm for the partition $\cals$ of $S$ that agrees with the
optimal partition $\cald$, i.e., $S_i = S \cap D_i$. (All lemmas
stated apply for this run of the algorithm, though we don't make this
explicit in the statement.) Let $(C_1, C_2, \dots, C_k)$ be the
clusters produced by the
algorithm at end of Step 4(c) on this run.  Let's begin with the
following simple observation.
\begin{lemma}
\label{lem:simple}
Suppose a node $u \in D_s$ is placed in cluster $C_r$ at the end of
Step 4(b) for $r\neq s$, $1 \le r,s \le k$. Then $\pval^\cald(u,r) \ge
\pval^\cald(u,s) - 2\beta = \val^\cald(u) - 2\beta$.
\end{lemma}
\proof Note that since $u \in D_s$, $\val^\cald(u) = \pval^\cald(u,s)$.
By the $\beta$-goodness of $S$ (recall Equation (\ref{eq:goodness})), $\pval^\cals(u,s) \ge \pval^\cald(u,s)  -
\beta$. Since we chose to place $u$ in $C_r$ instead of $C_s$, we must
have $\pval^\cals(u,r) \ge \pval^\cals(u,s)$. By the $\beta$-goodness
of $S$ again, we have $\pval^\cald(u,r) \ge \pval^\cals(u,r) -
\beta$. Combining these three inequalities gives us the claim of the
lemma. \QED

\smallskip
Define the set of nodes of low value in the optimal clustering $\cald$ as $T_\low \eqdef \{ u \mid \val^\cald(u) \le 1
- c_1/k^2 \}$. \red{The total number of disagreements is at least the number of disagreements induced by these low
valued nodes, therefore}
\begin{equation}
\label{eq:tlow-ub}
|T_\low| \le \frac{2 k^2 \disagr(\cald)}{(n-1) c_1}  =  \frac{2 k^2
  \gamma n^2}{(n-1) c_1}  \le \frac{4 k^2 \gamma n}{c_1} \ .
\end{equation}
The following key lemma
asserts that the large clusters produced in Step 4(c) are basically
correct.
\begin{lemma}
\label{lem:large-clus}
Suppose $\gamma \le \frac{c_1}{16 k^3}$. Let $\larg \subseteq
\{1,2,\dots,k\}$ be the set of large clusters as in Step 4(c) of the
algorithm. Then for each $i \in \larg$, $C_i - T_\low = D_i - T_\low$,
that is w.r.t nodes of large value, $C_i$ precisely agrees with the
optimal cluster $D_i$.
\end{lemma}
\proof Let $i \in \larg$ be arbitrary. We will first prove the
inclusion $C_i - T_\low \subseteq D_i - T_\low$. Suppose this is not
the case and there exists $u \in C_i - (D_i \cup T_\low)$. Let $u \in
D_j$ for some $j \neq i$. Since $u
\notin T_\low$, we have $\val^\cald(u) \ge 1 - c_1/k^2$, which implies
$\pval^\cald(u,j) \ge 1 - c_1/k^2$. By Lemma~\ref{lem:simple}, this
gives $\pval^\cals(u,i) \ge 1 -c_1/k^2 - 2\beta$. Therefore we have
\[ 2(1- c_1/k^2 - \beta) \le \pval^\cald(u,i) + \pval^\cald(u,j) \le 2 -
\frac{|D_i|+ |D_j| - 1}{n}
\]
where the last step follows from the simple but powerful observation
that each edge connecting $u$ to a vertex in $D_i \cup D_j$ is
correctly classified in exactly one of the two
placements of $u$ in the $i$'th and $j$'th clusters (when leaving
every other vertex as in clustering $\cald$).
We conclude that both
\begin{equation}
\label{eq:small-D}
|D_i|, |D_j| \le 2(\frac{c_1}{k^2}+\beta)n+1 \ .
\end{equation}
What we have shown is \red{t}hat if $u \in C_i - (D_i \cup T_\low)$, then $u \in D_j$ for some $j$ with $|D_j| \le
2(c_1/k^2+\beta)n+1$. It follows that $|C_i - (D_i \cup T_\low)| \le 2(c_1/k+\beta k)n+k$. Therefore,
\[ |D_i| \ge |C_i| - |T_\low| - 2(\frac{c_1}{k}+\beta k)n - k \ge
\frac{n}{2k} -
\frac{4 k^2 \gamma n}{c_1} - 2(\frac{c_1}{k}+\beta k)n - k >
2(\frac{c_1}{k^2}+\beta)n+1 \ , \] where the last step follows since
$\gamma \le \frac{c_1}{16  k^3}$, $k\ge 2$, $c_1 = 1/20$, and $\beta$
is tiny. This contradicts (\ref{eq:small-D}), and so we conclude $C_i
- T_\low \subseteq D_i -  T_\low$.

Now for the other inclusion $D_i - T_\low \subseteq C_i - T_\low$. If
a node $v \in D_i - (C_i \cup T_\low)$ is placed in $C_q$ for $q
\neq i$, then a similar argument to how we concluded
(\ref{eq:small-D}) establishes $|D_i| \le
2(\frac{c_1}{k^2}+\beta)n+1$, which is impossible since we have shown
$D_i \supseteq C_i - T_\low$, and hence $|D_i| \ge |C_i| - |T_\low|
\ge \frac{n}{2k} - \frac{4 k^2 \gamma n}{c_1} >
2(\frac{c_1}{k^2}+\beta)n+1$, where the last step follows using
$\gamma \le \frac{c_1}{16
  k^3}$ and $k\ge 2$ for the choice $c_1 = 1/20$. \QED

\smallskip The next lemma states that there is a clustering which is very close
to optimum which agrees exactly with our large clusters. This will
enable us to find a near-optimal clustering by recursing on the small
clusters to recluster them as needed, exactly as our algorithm does.
\begin{lemma}
\label{lem:large-clus-exact}
Assume $\gamma \le \frac{c_1}{16 k^3}$.
There exists a clustering $\calf$ that partitions $V$ as $V = F_1 \cup
F_2 \cup \cdots F_k$ that satisfies the following:
\begin{enumerate}
\item[(i)] $F_i = C_i$ for every $i \in \larg$
\item[(ii)] The number of disagreements of the clustering $\calf$ is
  at most $\disagr(\calf) \le \gamma n^2 \Bigl(1 + \frac{4 k^2}{c_1}  \bigl(\beta +  \frac{2 k^2 \gamma}{c_1}\bigr)\Bigr)$

\end{enumerate}
\end{lemma}
\proof Suppose $w \in T_\low$ is such that $w \in C_r$, $w \in D_s$
with $r \neq s$. Consider the clustering formed from $\cald$ by
performing the following in parallel for each $w \in T_\low$: If $w
\in C_r$ and $w \in D_s$ for some $r \neq s$, move $w$ to $D_r$.
Let $\calf = F_1 \cup \cdots \cup F_k$ be the resulting clustering.
By construction $F_i \cap T_\low = C_i \cap T_\low$ for all $i$, $1
\le i \le k$. Since we only move nodes in $T_\low$, clearly $F_i -
T_\low = D_i - T_\low$ for $1\ \le i \le k$. By
Lemma~\ref{lem:large-clus}, $C_i - T_\low = D_i - T_\low$ for $i \in
\larg$. Combining all these equalities we conclude that $F_i = C_i$
for each $i \in \larg$.

Now the only extra edges that the clustering $\calf$ can get wrong
compared to $\cald$ are those incident upon nodes in $T_\low$, and
therefore
\begin{equation}
\label{eq:hj-0}
\disagr(\calf) - \disagr(\cald) \le (n-1) \sum_{w \in T_\low}
  (\val^\cald(w) - \val^\calf(w))
\end{equation}
If a node $w$ belongs to the same cluster in $\calf$ and $\cald$
(i.e., we did not move it), then since no node outside $T_\low$ is
moved in obtaining $\calf$ from $\cald$, we have
\begin{equation}
\label{eq:hj-1}
\val^\calf(w) \ge \val^\cald(w) - |T_\low|/(n-1) \ .
\end{equation}
If we moved
a node $w \in T_\low$ from $D_s$ to $D_r$, then by
Lemma~\ref{lem:simple} we have $\pval^\cald(w,r) \ge \val^\cald(w) - 2\beta$.
Therefore for such a node $w$
\begin{equation}
\label{eq:hj-2}
\val^\calf(w) \ge \pval^\cald(w,r) -
|T_\low|/(n-1) \ge \val^\cald(w) - 2\beta - |T_\low|/(n-1) \ .
\end{equation}
Combining (\ref{eq:hj-0}), (\ref{eq:hj-1}) and (\ref{eq:hj-2}), we can
conclude $\disagr(\calf) - \disagr(\cald) \le (n-1) |T_\low| \bigl(2\beta +
\frac{|T_\low|}{n-1}\bigr)$. The claim now follows using the upper bound on
$|T_\low|$ from (\ref{eq:tlow-ub}) (and using $n^2/(n-1)^2 \le 2$).  \QED

\begin{lemma}
\label{lem:min-perf}
  If the optimal clustering $\cald$ has $\gamma n^2$ disagreements for
  $\gamma \le \frac{c_1}{16 k^3}$, then the clustering $\clus$ found by
  the algorithm makes at most $\gamma n^2 (1+\eps/10) \bigl(1 + 4 k^2
  \beta/c_1 + 8 k^4 \gamma/c_1^2\bigr)$ disagreements.
\end{lemma}
\proof We note that when restricted to the set of all edges except
those entirely within $W$, the set of agreements of the clustering
$\calc$ in Step 4(e) coincides precisely with that of $\calf$.  Let
$n_1$ be the number of disagreements of $\calf$ on edges that lie
within $W$ and let $n_2$ be the number of disagreements on all other
edges.  Since $W$ is clustered recursively, we have the number of
disagreements in $\calc$ is at most $n_2 + n_1 (1 +\eps/10) \le (n_1 +
n_2)(1+\eps/10)$. The claim follows from the bound on $n_1+n_2$ from
Lemma~\ref{lem:large-clus-exact}, Part (ii).  \QED

\begin{theorem}
  For every $\eps > 0$, algorithm {\bf MinDisAg}$(k,\eps)$ delivers a
  clustering with number of disagreements within a factor $(1+\eps)$
  of the optimum.
\end{theorem}
\proof Let $\opt =\gamma n^2 $ be the number of disagreements of an
optimal clustering. The solution $\clusmax$ returned by the
maximization algorithm has at most $\opt +
 \frac{\eps^2 c_1^2 n^2}{32 k^4}
= \gamma n^2 \Bigl(1
+ \frac{\eps^2 c_1^2}{32 k^4 \gamma}\Bigr)$ disagreements. The solution $\clus$ has at most
$\gamma n^2 (1+\eps/10) \bigl(1 + 4 k^2\beta/c_1 + 8 k^4
\gamma/c_1^2)\bigr)$ disagreements. If $\gamma > \frac{\eps c_1^2}{32
  k^4}$, the former is within $(1+\eps)$ of the optimal. If $\gamma
\le \frac{\eps c_1^2}{32 k^4}$ (which also satisfies the requirement
$\gamma \le c_1/16k^3$ we had in Lemma~\ref{lem:min-perf}), the latter clustering $\clus$ achieves approximation
ratio $(1+\eps/10)(1+\eps/2) \le (1+\eps)$ (recall that $\beta \le
\frac{\eps c_1}{16 k^2}$).  Thus the better of these two solutions is
always an $(1+\eps)$ approximation.  \QED

\smallskip
\red{To conclude Theorem~\ref{thm:min-ptas}, we examine the running time of {\bf MinDisAg}. Step 4 will be run for
$k^{|S|}=n^{O(k^{4}/ \epsilon^{2})}$ iterations. During each iteration, the placement of vertices is done in $O(n\log
n)$. Finally, observe that there is always at least one large cluster, therefore the recursive call is always done on
strictly less clusters. It follows that the running time of {\bf MinDisAg}$(k,\eps)$ can be described from the
recurrence $T(k,\epsilon)\leq n^{O(k^{4}/ \epsilon^{2})}(n\log n + \cdot T(k-1,\epsilon/10))$ from which we derive that
the total running time is bounded by $n^{O(100^{k}/ \epsilon^{2})}\log n$.}

\vspace{-1ex}
\section{Complexity on general graphs}
\label{sec:general-ghs}
So far, we have discussed the \maxagr{k} and \mindisagr{k} problems
on complete graphs. In this section, we note some results on the
complexity of these graphs when the graph can be arbitrary. As we will
see, the problems become much harder in this case.
\begin{theorem}
\label{thm:gen-maxagr} There is a polynomial time factor $0.878$ approximation algorithm for \maxagr{2} on general
graphs.  For every $k \ge 3$, there is a polynomial time factor $0.7666$ approximation algorithm for \maxagr{k} on
general graphs.
\end{theorem}
\proof The bound for $2$-clusters case follows from the
Goemans-Williamson algorithm for Max CUT modified in the obvious way
to account for the positive edges. The bound for $k \ge 3$ is obtained
by Swamy~\cite{swamy} who also notes that slightly better bounds are
possible for $3 \le k \le 5$. \QED

\smallskip\noindent
We note that in light of the recent hardness result for Max
CUT~\cite{KKMO}, the above guarantee for \maxagr{2} is likely the best
possible.
\begin{theorem}
  There is a polynomial time \red{$O(\sqrt{\log n})$} approximation
  algorithm for \mindisagr{2} on general graphs. For $k \ge 3$,
  \mindisagr{k} on general graphs cannot be approximated within any
  finite factor.
\end{theorem}
\proof The bound for $2$-clustering follows by the simple observation that \mindisagr{2} on general graphs reduces to
Min 2CNF Deletion, i.e., given an instance of 2SAT, determining the minimum number of clauses that have to be deleted
to make it satisfiable. \red{The latter problem admits an $O(\sqrt{\log n})$ approximation algorithm}~\cite{ACMM}. The
result on \mindisagr{k} for $k \ge 3$ follows by a reduction from $k$-coloring. When $k \ge 3$, it is NP-hard to tell
if a graph is $k$-colorable, and thus even given an instance of \mindisagr{k} with only negative edges, it is NP-hard
to determine if the optimum number of disagreements is zero or positive. \QED

\bibliographystyle{plain}
\bibliography{corrclus}

\end{document}